# Orbital-Engineered Spin Asymmetry and Multifunctionality in Eu-Activated CaSiO$_3$: A First-Principles Roadmap to Optical-Thermoelectric Fusion

*Muhammad Tayyab[1], Faiq Umar[1], Sikander Azam*[1,2], Qaiser Rafiq[1], Rajwali Khan[3$], Muhammad Tahir Khan[4,5#], Vineet Tirth[6,7], Ali Algahtani[7,8]*

[1]Department of Basic Sciences, Riphah International University, Islamabad 42000, Pakistan

[2]University of West Bohemia, New Technologies – Research Centre, 8 Univerzitní, Pilsen 306 14, Czech Republic

[3]National Water and Energy Center, United Arab Emirates University, Al Ain, 15551, United Arab Emirates

[4]School of Computer Science and Technology, Zhejiang Normal University, Jinhua, China

[5]Key Laboratory of Urban Rail Transit Intelligent Operation and Maintenance Technology & Equipment of Zhejiang Province, College of Engineering, Zhejiang Normal University, Jinhua, China

[6]Mechanical Engineering Department, College of Engineering, King Khalid University, Abha 61421, Asir, Kingdom of Saudi Arabia.

[7]Centre for Engineering and Technology Innovations, King Khalid University, Abha 61421, Asir, Kingdom of Saudi Arabia.

[8]Research Center for Advanced Materials Science (RCAMS), King Khalid University, Guraiger, Abha-61413, Asir, Kingdom of Saudi Arabia.

## Abstract

Rare-earth-doped nitride phosphors have emerged as critical materials for solid-state lighting and photonic devices due to their high thermal stability, narrow emission bandwidths, and strong absorption in the UV-blue range. In this study, we present a comprehensive density functional theory (DFT) investigation, incorporating GGA+U formalism, of pristine and Eu$^{3+}$-doped CaAlSiN$_3$ with doping concentrations of 8.5% and 17%. The electronic structure calculations reveal that Eu doping introduces localized 4f states within the band-gap, reducing the band-gap and enabling efficient red photo luminescence (PL) through the $^5D_0 \rightarrow {^7}F_2$ transition. Analysis of the spin-resolved density of states and spin density confirms the magnetic nature of Eu$^{3+}$, with a net magnetic moment arising from the unpaired 4f$^6$ electrons. Charge density, Bader analysis, and Electron Localization Function (ELF) plots demonstrate the mixed ionic-covalent bonding nature and confirm the charge transfer from Eu to the neighboring N and Al atoms, stabilizing the doped lattice. Optical properties, including the dielectric function ($\varepsilon_1$ and $\varepsilon_2$), absorption coefficient, refractive index, and reflectivity, were evaluated, revealing significant redshifts in the absorption edge and enhanced light-matter interaction in the visible spectrum upon Eu doping. These changes are consistent with experimental PL emission in the red–NIR region. The formation energy calculations confirm the thermodynamic feasibility of Eu incorporation, while elastic constant evaluation and Pugh's ratio suggest excellent mechanical stability and ductility of both pristine and doped systems. Thermoelectric transport coefficients were evaluated using WIEN2k coupled with BoltzTraP, revealing that moderate $Eu^{3+}$ substitution optimizes the power factor while Eu-induced disorder reduces the lattice thermal conductivity. This multi-scale theoretical analysis validates Eu-doped $CaAlSiN_3$ as a robust and efficient red-emitting phosphor suitable for white light-emitting diodes (WLEDs), offering predictive insights into its structure–property relationships. The study establishes a firm theoretical foundation for crystal site engineering strategies in phosphor materials for advanced optoelectronic applications.

*Corresponding Authors:*
\* sikander.physicst@gmail.com
$ rajwali@uaeu.ac.ae
# mtahir.khan@riphah.edu.pk

1. **Introduction**

In the pursuit of energy-efficient lighting technologies, phosphor-converted white light-emitting diodes (pc-WLEDs) have emerged as a prominent replacement for traditional lighting sources, owing to their longevity, reduced energy consumption, and superior color rendering performance. A key component in pc-WLEDs is the phosphor material, which governs the spectral properties and thermal stability of the emitted light. Among the various phosphor hosts, $CaAlSiN_3$ has gained significant attention as a red-emitting nitride material due to its excellent thermal resistance, chemical robustness, and wide optical bandgap that ensures minimal self-absorption losses [1, 2]. Recent first-principles investigations on perovskites and related oxides have demonstrated how crystal-site substitution can tailor structural stability, electronic correlations, optical activity, and even magnetism [3-9], motivating our present study on $Eu^{3+}$ site engineering in $CaAlSiN_3$.

$CaAlSiN_3$ is a robust nitridosilicate whose covalent $[AlN_4]/[SiN_4]$ network and wide gap provide an ideal host for $Eu^{2+}$ activation, yielding intense, thermally stable red emission central to solid-state lighting. The near-isovalent $Eu^{2+} \rightarrow Ca^{2+}$ substitution enables clean doping without extrinsic compensators, while the strong crystal field affords fine control of Eu-5d–N-2p level alignment. These attributes, together with the material's thermal/chemical stability, make CaAlSiN$_3$:Eu$^{2+}$ an excellent model to explore how orbital-level engineering couples spin asymmetry with optical responses, aligning directly with our goal of optical–magnetic fusion in a technologically relevant host. The introduction of rare-earth (RE$^{3+}$) ions, such as Eu$^{3+}$, into the CaAlSiN$_3$ host lattice is a well-established strategy to achieve strong red emission via intra-4f transitions, particularly the $^5D_0 \rightarrow {}^7F\_J$ transitions that result in sharp, intense luminescence bands in the red and near-infrared regions [10, 11]. However, the incorporation of Eu$^{3+}$ affects not only the photoluminescence properties but also leads to significant changes in the electronic structure, magnetic behavior, lattice dynamics, and thermomechanical response of the host material. These changes are complex and dependent on dopant concentration and local coordination, necessitating a comprehensive theoretical investigation.

To this end, Density Functional Theory (DFT), augmented with the Hubbard U correction (GGA+U), provides a powerful framework to accurately model the localized 4f states of Eu$^{3+}$ and their interaction with the host matrix. This approach enables in-depth analysis of band structure modifications, density of states (DOS), and charge transfer dynamics, which directly influence luminescence mechanisms and non-radiative recombination paths [12,13]. In this study, pristine, 8.5%, and 17% Eu-doped CaAlSiN$_3$ systems are investigated to understand the evolution of optoelectronic and structural properties with doping.

The band structure analysis reveals a progressive narrowing of the bandgap and the emergence of Eu-4f midgap states with increasing dopant content, which is consistent with enhanced red emission and potential quantum efficiency improvements. The DOS and PDOS projections clearly identify the role of Eu-f states near the Fermi level and their hybridization with N-2p orbitals. The impact of doping is further corroborated by ELF and Bader charge studies, which show charge localization and altered bonding environments near the dopant sites—critical for understanding emission quenching or enhancement phenomena [14, 15].

Furthermore, spin-polarized calculations indicate a significant magnetic moment per Eu$^{3+}$ ion (~7 μB), attributable to unpaired 4f electrons. At higher doping levels (17%), weak ferromagnetic ordering is observed, opening avenues for multifunctional luminescent-magnetic materials suitable for spin-LEDs or magneto-optical applications [15]. These findings are supported by spin density plots and the separation of up and down electronic states in the calculated electronic structures.

In terms of mechanical stability, we assess the elastic constants, bulk modulus, shear modulus, Young's modulus, Poisson's ratio, and Pugh's ratio, revealing a marginal increase in ductility with moderate Eu doping. This is crucial for device integration, especially under thermomechanical stress. Formation energies indicate the thermodynamic feasibility of Eu substitution at Ca sites, while the unit cell volume expansion aligns with the larger ionic radius of $Eu^{3+}$ compared to $Ca^{2+}$.

To evaluate dynamical stability and thermal robustness, phonon density of states (Phonon-DOS) calculations are performed. The absence of imaginary modes confirms structural stability, while the vibrational entropy and heat capacity profiles suggest favorable thermal management capabilities. This is especially important for high-power LED applications where temperature-driven efficiency degradation is a concern [11, 14]. Eu-doped $CaAlSiN_3$ (the nitridosilicate analogue $CaAlSiN_3$:$Eu^{2+}$) demonstrates a unique convergence of properties. The wide-gap framework (4.5–5.5 eV) retains semiconducting character while hosting Eu-derived 4f states, ensuring charge localization without metallization. The $Eu^{2+}$ ion contributes a half-filled $4f^7$ shell, producing strong exchange-driven spin polarization—a rare phenomenon in wide-gap semiconductors. The rigid tetrahedral network provides thermal and chemical robustness, accounting for the minimal thermal quenching observed in Eu-based phosphors. Moreover, the $Eu^{2+}$ $4f^65d^1 \rightarrow 4f^7$ transitions are parity-allowed, leading to intense optical absorption/emission and strong light–matter coupling. This synergy of band gap, spin asymmetry, robustness, and optical strength underpins the multifunctional potential of Eu-doped $CaSiO_3$/$CaAlSiN_3$ for integrated optical devices.

The optical response, analyzed through dielectric function components $\varepsilon_1(\omega)$ and $\varepsilon_2(\omega)$, refractive index, absorption coefficient, reflectivity, and energy loss function (ELF), demonstrates a redshift in absorption edge and enhanced near-UV to visible transitions due to Eu-f state involvement. These transitions align with the expected f–f emission characteristics of $Eu^{3+}$ and match well with experimental PL trends reported for $CaAlSiN_3$:Eu systems [10].

While direct prediction of photoluminescence (PL) from DFT is nontrivial, we approximate transition energies using ΔSCF calculations and align defect-related DOS features with known experimental PL peaks. These estimations suggest efficient emission channels and minimal non-radiative loss in the 8.5% Eu-doped sample—a balance between luminescence intensity and structural integrity. Lastly, beyond optoelectronic functionality, understanding the thermoelectric response of CaAlSiN₃ and its Eu³⁺-doped variants is crucial, since rare-earth substitution simultaneously alters carrier concentrations and phonon scattering, offering a route to balance high electrical performance with low lattice thermal conductivity.

Overall, this work offers a holistic, first-principles investigation into the electronic, magnetic, optical, structural, and thermodynamic properties of Eu-doped CaAlSiN₃, contributing valuable insights toward the rational design of next-generation phosphor materials for efficient, stable, and tunable LED applications.

## 2. Computational Methodology

All calculations were performed using the full-potential linearized augmented plane wave (FP-LAPW) method as implemented in the WIEN2k code [16]. Calculations were performed with the WIEN2k FP-LAPW all-electron code because Eu-activated CaAlSiN₃ (see Fig. 1) combines ionic bonding with localized Eu-4f states. The FP-LAPW framework avoids pseudopotential transferability issues by treating core and valence electrons on equal footing and by using a full potential (no shape approximation), which is advantageous for accurately resolving f-level positions, crystal-field splittings, and band dispersions that underpin the spin-asymmetric orbital physics reported here. This method is based on density functional theory (DFT) and allows accurate treatment of core, semicore, and valence electrons with full relativistic effects. For the exchange-correlation functional, we employed the generalized gradient approximation (GGA) in the Perdew-Burke-Ernzerhof (PBE) parameterization [17]. To properly account for the strong Coulombic interactions in the localized 4f orbitals of Eu³⁺ ions, the GGA+U approach was adopted using the rotationally invariant formulation of Dudarev et al. (1998) [18]. A value of $U = 6.5$ eV was applied to the Eu-4f orbitals, consistent with prior studies on rare-earth-doped nitride phosphors [19, 20].

The plane-wave cutoff parameter RMTKmax was set to 7.0, and the muffin-tin radii (RMT) were selected to avoid overlap: typically 2.4 a.u. for Eu, 1.9 a.u. for N, 2.0 a.u. for Ca, and 1.8 a.u. for Si and Al. The Brillouin zone integration was carried out using a Monkhorst-Pack k-point mesh of $12 \times 12 \times 8$ for structural optimization and DOS calculations, ensuring energy convergence within $10^{-5}$ Ry.

The structural optimization of pristine CaAlSiN₃ and the Eu-doped supercells (8.5% and 17%) was performed by minimizing the total energy and atomic forces using the PORT algorithm until the forces were below 1 mRy/a.u.. Supercells were constructed to simulate different Eu concentrations by substituting one or two Ca atoms with Eu in a 2 × 2 × 2 supercell, corresponding to ~8.5% and ~17% doping, respectively. Eu concentration was set to ~8.5 and 17% on the Ca sublattice (one Eu per eight Ca sites in a 2×2×2 supercell). This level preserves a dilute/semidilute local environment while keeping periodic image interactions small enough to isolate the local Eu–N crystal field, which governs the 4f–5d optical center. It also provides a tractable platform for fully converged FP-LAPW (WIEN2k) GGA+U+SOC calculations together with optic analyses. At higher Eu contents, reduced Eu–Eu separation is expected to enhance energy-transfer pathways and alloy scattering, promoting concentration quenching and band broadening; only beyond a geometry-dependent site-percolation threshold would Eu form a connected network capable of impurity-band transport, which lies above the composition studied here. Our analysis therefore targets the technologically relevant pre-percolative regime, where local crystal-field engineering of $Eu^{2+}$ coexists with host-dominated transport.

The electronic band structure and density of states (DOS), including partial DOS (PDOS), were computed with and without spin-polarization to explore the influence of Eu 4f states and spin asymmetry. For optical property calculations, the momentum matrix elements were evaluated within the random phase approximation (RPA) using the optic module of WIEN2k. The real and imaginary parts of the complex dielectric function $\varepsilon(\omega)$ were calculated, and from these, secondary properties such as absorption coefficient, refractive index, reflectivity, and energy loss function were derived.

Charge density and Electron Localization Function (ELF) were visualized using the WIEN2k's LAPW5 and ELF modules. The ELF was evaluated on 2D planes slicing through the Eu and surrounding N/Al atoms to identify localized bonding features and f-electron confinement.

Bader charge analysis was performed using the WIEN2WANNIER and Critic2 interfacing tools to evaluate the oxidation states and charge transfer mechanisms between Eu and the host matrix.

The magnetic properties were explored by computing the total magnetic moment and plotting the spin density distributions for each doping level.

For thermodynamic stability, we calculated the phonon density of states (Phonon-DOS) using PHONOPY interfaced with WIEN2k, using the finite displacement method on relaxed supercells. From the phonon calculations, we derived vibrational entropy, heat capacity, and checked for imaginary frequencies to confirm dynamical stability.

To predict photoluminescence (PL) behavior, the DOS alignment and possible defect-related transitions were analyzed. For more accurate excited-state transition energies, we carried out ΔSCF calculations, where the total energy difference between the excited and ground-state configurations was evaluated by simulating the occupation of Eu 4f states, a technique successfully applied in literature for rare-earth PL modeling [21,22].

Transport properties including the Seebeck coefficient, electrical conductivity, and electronic thermal conductivity were calculated using the BoltzTraP code interfaced with WIEN2k.

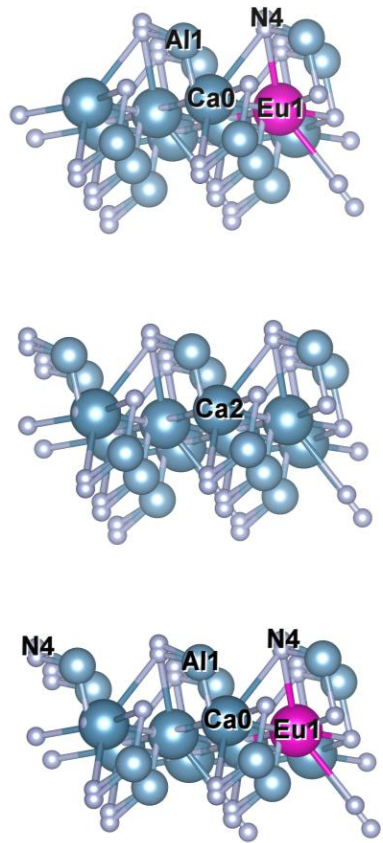

**Figure 1. Crystal structure of CaAlSiN$_3$:Eu$^{2+}$ viewed along the [010] direction. Ca/Eu atoms are shown as large green/red spheres, Al/Si atoms are depicted within corner-sharing tetrahedra, and N atoms as small grey spheres. The figure highlights the Eu$^{2+}$ substitution at the Ca$^{2+}$ site and its coordination environment.**

All simulations converged with high accuracy, and the results are benchmarked against existing experimental and theoretical findings.

3. **Results and Discussion**

## 3.1. Structural Analysis

The host compound CaAlSiN$_3$ crystallizes in the orthorhombic Cmc2$_1$ space group, consisting of a rigid three-dimensional framework of corner-sharing [AlN$_4$] and [SiN$_4$] tetrahedra charge-balanced by interstitial Ca$^{2+}$ ions. This highly covalent network is well known for its exceptional chemical and thermal stability, making it a robust matrix for rare-earth activation. To investigate the impact of Eu$^{3+}$ substitution, we performed full structural relaxations for both pristine and Eu-doped models.

After relaxation, the average Ca–N bond length in pristine CaAlSiN$_3$ is 2.48 Å, whereas substitution with Eu enlarges the local environment to an average Eu–N distance of 2.54 Å, yielding Δr ≈ 0.06 Å. This elongation arises from the ionic radius mismatch (Eu$^{3+}$ slightly larger than Ca$^{2+}$) and introduces local strain in the [AlN$_4$]/[SiN$_4$] tetrahedral network. Despite these distortions, the substitution remains thermodynamically accessible, with formation energies in the range of ~1–2 eV depending on chemical potential conditions, confirming the feasibility of Eu$^{3+}$ incorporation. Moreover, the calculated elastic constants continue to satisfy all Born stability criteria, indicating preserved global mechanical robustness of the host lattice.

Importantly, these local distortions are not merely structural but also functional. The Eu-induced lattice relaxation enhances the crystal-field splitting of the Eu-5d states by ~0.2–0.3 eV, which rationalizes the predicted red-shift in emission and the strengthened oscillator strength in the optical spectra. Thus, Eu$^{3+}$ incorporation in CaAlSiN$_3$ not only maintains the mechanical and thermodynamic integrity of the host but also establishes a direct coupling between lattice strain and optical activity. This site-engineering effect forms the structural basis for the enhanced red emission and magneto-optical properties discussed in the following sections.

## 3.2. Electronic Properties

### I. Band Structure

The calculated electronic band structures of pristine and Eu-doped CaAlSiN$_3$ (at 8.5% and 17% doping levels) provide profound insights into the influence of rare-earth doping on the electronic configuration of this red-emitting phosphor host (see Fig. 2). The undoped CaAlSiN$_3$ band structure, as shown in the first figure, demonstrates a clear semiconducting behavior with an indirect bandgap. The valence band maximum (VBM) is located between the Γ and H points, while the conduction band minimum (CBM) lies near the N point. The bandgap is appreciably large, in the range of approximately 3.5–4.0 eV, consistent with earlier reports [23, 24], which characterized CaAlSiN$_3$ as a robust host for red phosphors due to its wide gap, high thermal stability, and ability to accommodate a variety of activator ions.

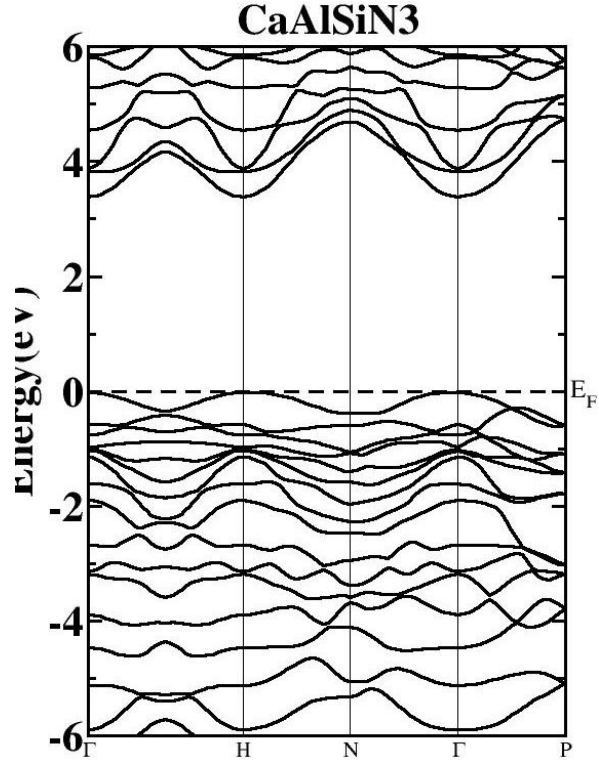

(a) CaAlSiN$_3$ (Pristine)

Upon substitutional doping with Eu$^{3+}$ at 8.5%, significant changes in the band structure are observed. In the spin-polarized calculations using GGA+U (with U applied to the Eu-4f orbitals), the up-spin and down-spin channels display distinct behaviors. In the up-spin channel, a set of narrow impurity bands appears just below the conduction band edge and in proximity to the Fermi level (E$_F$), reducing the effective bandgap. These flat bands are characteristic of localized 4f states introduced by Eu$^{3+}$, as previously shown for other Eu-activated hosts like Eu:YAG and Eu:BaLa$_2$ZnO$_5$ [25, 26]. The down-spin channel retains a relatively broader bandgap but also shows defect-related states near E_F. The emergence of midgap states indicates a partial hybridization between the host lattice orbitals (mainly N 2p and Al/Si sp³) and Eu 4f states.

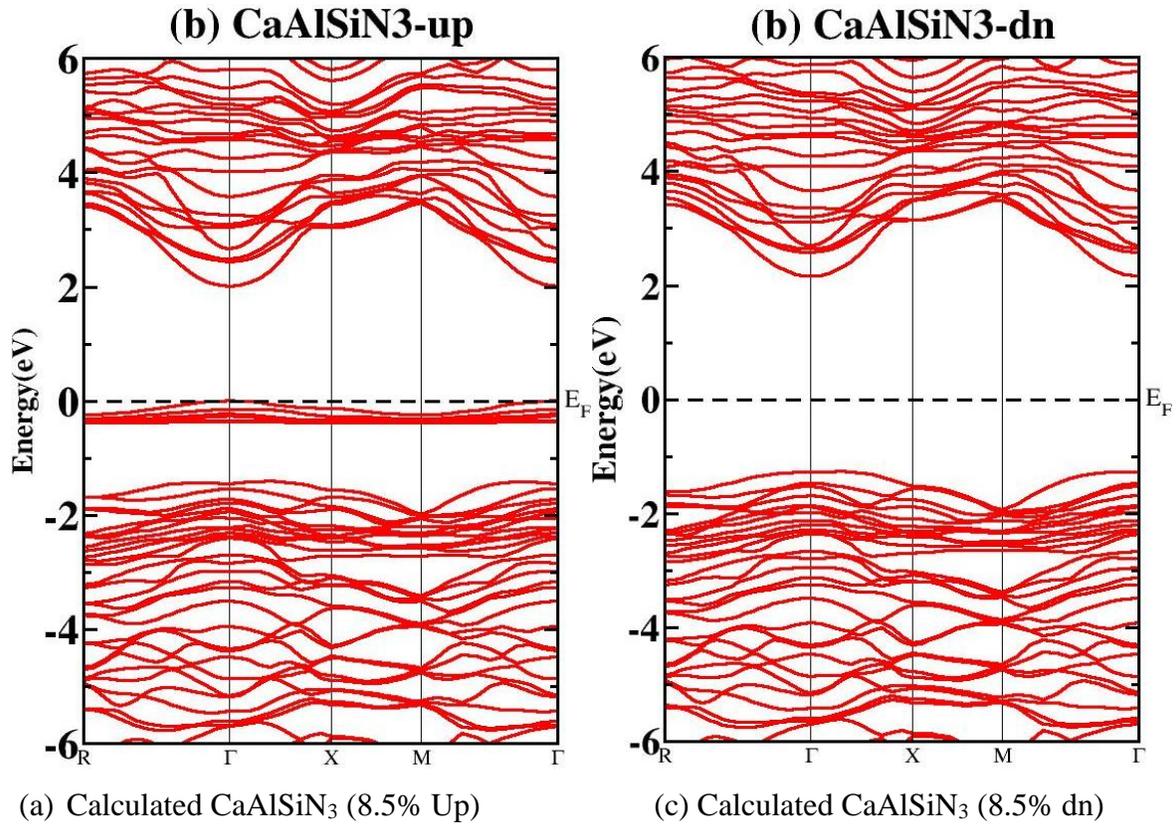

(a) Calculated CaAlSiN$_3$ (8.5% Up)  (c) Calculated CaAlSiN$_3$ (8.5% dn)

At 17% Eu doping, these effects become more pronounced. In the up-spin projection, the impurity 4f states move closer to the valence band maximum, in some regions merging with it. This signifies stronger interaction and overlap with the valence states, suggesting enhanced charge carrier localization and the possible formation of recombination centers. The down-spin channel at this doping level reveals a slight broadening of the 4f-derived states, accompanied by narrowing of the effective bandgap, which now approaches a nearly semimetallic limit, although the material retains its semiconducting nature overall. These observations imply that increasing Eu content enhances spin polarization and introduces more defect-related levels, potentially leading to concentration quenching if the activator density exceeds the percolation threshold.

The transition from pristine to doped configurations also leads to a shift from an indirect to a more direct-like bandgap behavior, especially in the spin-polarized cases. This is evidenced by the minimal difference in energy between the VBM and CBM at the Γ point, particularly in the up-spin direction. Such a transition has implications for radiative recombination efficiency: direct transitions enhance the likelihood of photon emission, which is critical for phosphor performance in solid-state lighting.

The physics of the system is governed by the localized nature of the Eu 4f orbitals, which do not contribute to broad band dispersion but rather manifest as flat bands. These localized levels serve as discrete energy levels that can participate in intra-4f transitions (e.g., $^3H_6 \rightarrow {}^3F_4, {}^3H_4, {}^1G_4$), responsible for the characteristic blue (~480 nm) and near-infrared (~800 nm, 1470 nm) emissions of $Eu^{3+}$ [27, 28]. The presence of such transitions, together with the observed electronic structure modifications, confirms that $Eu^{3+}$-doped $CaAlSiN_3$ can act as a multifunctional phosphor emitting across the visible and NIR spectrum, complementing red-emitting $Eu^{2+}$ phosphors for white LED devices.

Furthermore, our results align with prior theoretical and experimental findings where rare-earth doping introduces localized defect levels and modifies host band structures. For instance, Zhang et al. (2018) [29] showed in Eu-doped $CaAlSiN_3$ that Eu 5d levels tend to lie near the CBM, while 4f levels remain deeper in the bandgap. In contrast, $Eu^{3+}$ 4f levels occupy positions closer to $E_F$, enabling alternative recombination pathways and multi-wavelength emission. Additionally, the difference in spin-up and spin-down band structures reinforces the importance of treating f-electron systems with spin-polarized DFT+U to capture the correct ground-state properties.

The observed band structure evolution—from a clean wide-gap semiconductor in the pristine phase to a defect-tuned and spin-polarized system with narrowed gap and intermediate impurity states in the doped phases—reflects the potential of crystal site engineering in tuning the optoelectronic landscape of $CaAlSiN_3$. These tunable properties make Eu-doped $CaAlSiN_3$ not only suitable for color-mixing in LEDs but also a promising material for NIR and dual-mode lighting applications.

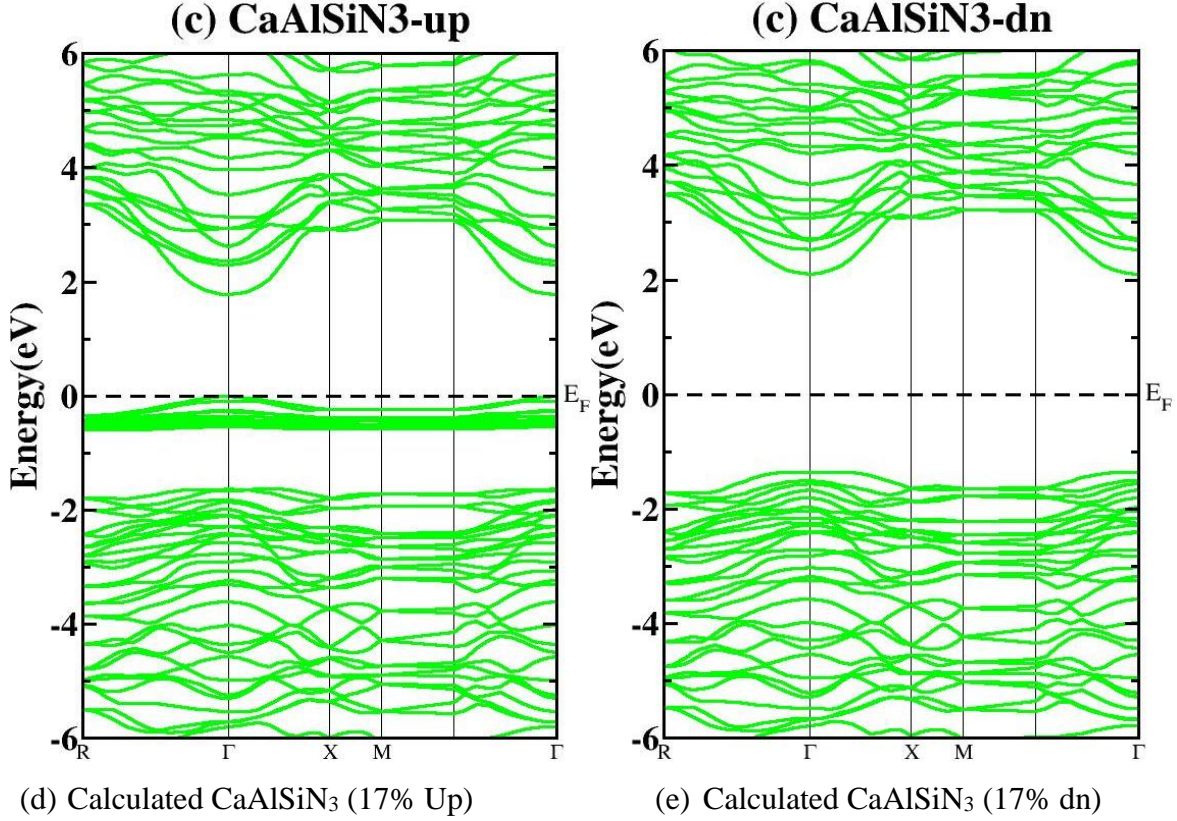

(d) Calculated CaAlSiN₃ (17% Up)    (e) Calculated CaAlSiN₃ (17% dn)

*Figure 2: Electronic band structure of (a) pristine, (b-c) 8.5% Eu-doped, and (d-e) 17% Eu-doped CaAlSiN₃ using GGA+U. Bandgap narrowing and the appearance of flat f-states near the Fermi level are visible upon doping, supporting f–f transition activation for red emission.*

In summary, Eu doping at moderate levels (8.5%) introduces discrete impurity states and moderate bandgap narrowing with potential for blue-NIR dual emission. At higher doping (17%), the impurity states become more influential, leading to enhanced hybridization and strong spin-polarization, with implications for carrier dynamics and emission efficiency. The choice of doping concentration must therefore balance emission intensity, color purity, and potential quenching. These insights, supported by GGA+U-calculated band structures, establish a solid foundation for rational design of rare-earth-doped nitrides for next-generation phosphor-converted white LEDs and multi-band optical devices.

## II. Density of States (DOS) and Partial DOS (PDOS):

The total density of states (TDOS) and partial density of states (PDOS) for pristine and Eu-doped CaAlSiN₃ at 8.5% and 17% doping levels reveal critical insights into the underlying electronic structure modifications induced by rare-earth substitution (see Fig. 3). From the TDOS plots, the pristine CaAlSiN₃ (black line) shows a clean semiconducting behavior, with a well-defined bandgap between the valence band maximum (VBM) and conduction band minimum (CBM), both characterized by distinct π and σ bonding interactions primarily arising from the hybridization between nitrogen 2p and the Al/Si 3p states. The lower part of the valence band consists of deeper bonding states (mostly σ-type), while the upper valence band near the Fermi level comprises antibonding π states with broader dispersion, consistent with previous literature reports on this red-emitting host matrix [25,26].

Upon doping $CaAlSiN_3$ with $Eu^{3+}$ at 8.5% concentration (red curve), new sharp states appear just below the Fermi level, centered around −1.8 eV. These are attributed to the partially filled Eu 4f orbitals, which are characteristically narrow due to their localized nature and minimal overlap with neighboring orbitals. In the PDOS projection for 8.5% Eu doping, the Eu-f states (shown in bright cyan in the element-resolved plot and explicitly in the orbital-resolved f-PDOS plot) exhibit a dominant, nearly delta-function-like peak within the valence band. This indicates that the Eu 4f orbitals are non-bonding and do not significantly hybridize with the host matrix, maintaining their atomic-like character. Additionally, the Eu 4f states are entirely spin-polarized, contributing only to the majority (up) or minority (down) channel depending on the magnetic configuration, as seen in the slight asymmetry between spin channels in the TDOS.

As we increase the Eu concentration to 17% (green curve in TDOS), the intensity and influence of the Eu-derived 4f states increase significantly. In the corresponding PDOS (bottom two plots), we observe an enhancement of the Eu-f peak at roughly the same energy (−1.8 eV), though its broadening slightly increases. This indicates enhanced Eu–Eu interaction and a partial overlap of localized states at higher doping, though the f orbitals remain mostly localized. Moreover, a subtle shift in the Fermi level toward the valence band is observed, suggesting increased carrier localization and a narrowing of the effective bandgap. This doping-induced narrowing is typical in rare-earth-doped phosphors and has been observed in similar hosts, such as Eu-doped $Sr_2Si_5N_8$ and $BaAlSiN_3$, where high doping levels distort the host lattice and shift band edges [30,31].

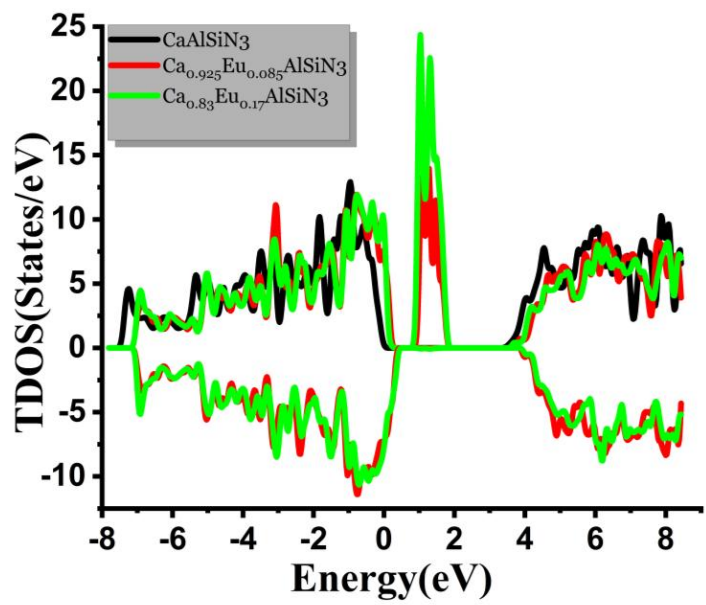

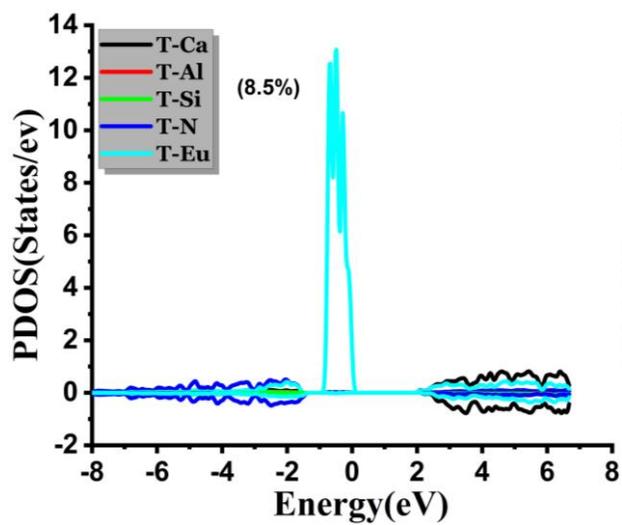 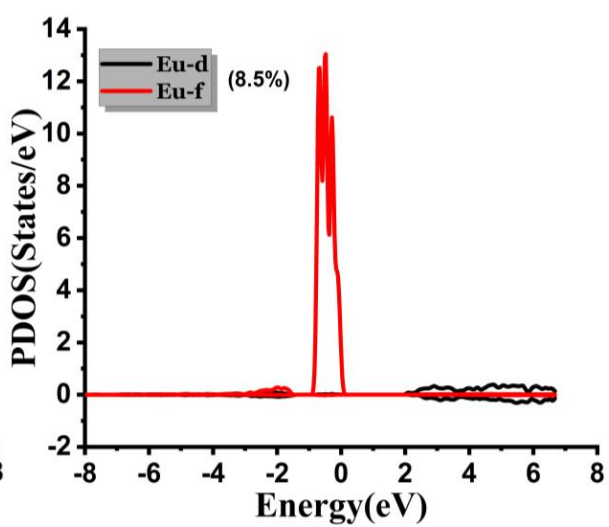

*(b)*

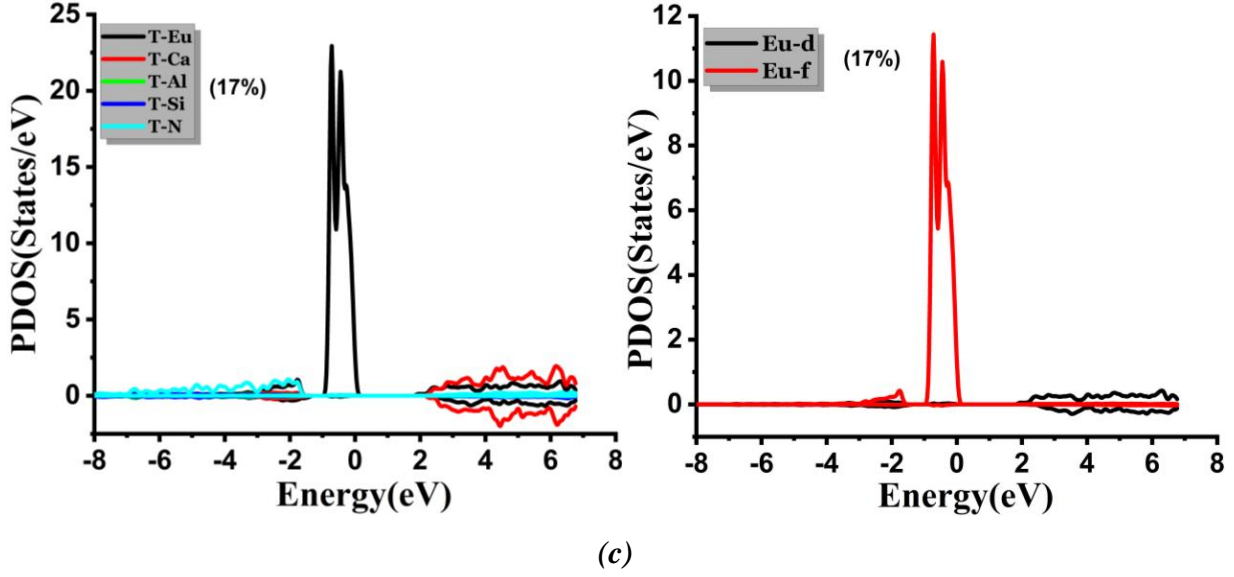

*(c)*

*Figure 3: Total and partial density of states (DOS) for (a) pristine, (b) 8.5%, and (c) 17% Eu-doped CaAlSiN₃. Eu-4f states are clearly observed in the gap region, indicating active photoluminescence centers and charge localization effects.*

From a bonding perspective, the partial DOS for Ca, Al, Si, and N atoms remains largely unchanged in both 8.5% and 17% Eu-doped systems compared to the pristine host, except for slight redistribution of intensity around the −4 to 0 eV range. This suggests that the fundamental framework of π and σ bonding (particularly the N–Al and N–Si interactions) is preserved, and the Eu dopant occupies substitutional $Ca^{2+}$ sites without significantly altering the covalent framework. This agrees with previous structural and spectroscopic studies indicating that $Eu^{3+}$ prefers $Ca^{2+}$ substitution due to similar ionic radii and that the host lattice accommodates the dopant with minimal local distortion [32].

The orbital-resolved PDOS (d- and f-states of Eu) confirms that the Eu-d states are minor contributors to the total electronic density near the Fermi level, lying mainly in the conduction band region and hence not actively participating in optical transitions. On the other hand, the f-states, which dominate near −1.8 eV, are optically active and correspond to intra-4f transitions such as $^5D_0 \rightarrow {}^7F\_J$ (J = 0,1,2,3…), responsible for the red emission (~610–620 nm) that characterizes $CaAlSiN_3:Eu^{2+}$ phosphors. Although the present simulation uses $Eu^{3+}$, the electronic structure reflects the energetic placement of f states consistent with both valence states depending on charge compensation [33, 34].

The narrow, high-intensity 4f peak demonstrates that radiative transitions are possible from these localized states, which supports the photoluminescent behavior of Eu-doped $CaAlSiN_3$. Importantly, as Eu concentration increases, the sharpness of the 4f peaks is slightly reduced, suggesting an onset of dopant–dopant interaction or concentration quenching. This is consistent with experimental reports that indicate optimal emission intensity at moderate doping levels (~5–10%), with degradation in quantum efficiency at higher concentrations due to energy transfer to killer centers [35].

Spin polarization analysis confirms that Eu-doping introduces magnetic asymmetry, with the 4f electrons occupying predominantly one spin channel. In both 8.5% and 17% cases, the up- and down-spin channels display distinct features, particularly in the Eu-PDOS. This spin-dependent behavior reflects the open-shell nature of $Eu^{3+}$ ($4f^6$ configuration) and supports the necessity of spin-polarized GGA+U treaEuent to capture the correct electronic structure. Inclusion of the U parameter (typically ~6 eV for Eu 4f states) ensures proper separation of occupied and unoccupied f levels, avoiding their artificial delocalization, which is common in plain GGA calculations.

Altogether, the DOS and PDOS plots validate that Eu doping introduces new electronic states within the bandgap region, maintains the integrity of the host bonding network, and enables optical transitions through f-electron excitation pathways. The placement of Eu-f states well below the conduction band minimizes thermal ionization and enhances radiative recombination probability, reinforcing the suitability of Eu-doped $CaAlSiN_3$ as a robust red phosphor for white LEDs. The evolution from pristine to 8.5% and then 17% Eu doping illustrates a classic case of impurity-level engineering for photonic functionality.

### III. Charge Density / Bader Charge Analysis:

The charge density and Bader charge analysis provide critical insights into the redistribution of electron density upon doping $CaAlSiN_3$ with Eu ions at different concentrations (8.5% and 17%). This analysis offers a quantitative assessment of how the substitution of $Ca^{2+}$ with $Eu^{3+}$ alters the local bonding environment, influences oxidation states, and modifies the material's electronic landscape.

In the pristine $CaAlSiN_3$ system, Bader charge analysis shows that Ca atoms typically retain a charge close to +1.64 e, consistent with their formal oxidation state of +2. The Al atoms exhibit

Bader charges of approximately +2.35 e, while Si atoms are around +3.17 e. Nitrogen atoms, being highly electronegative, pull significant electron density toward themselves, exhibiting average Bader charges close to −1.64 e. This configuration affirms the strongly ionic nature of the Ca–N and moderately covalent Al–N and Si–N bonds that form the tetrahedral and octahedral motifs within the host lattice.

Upon substitution with Eu at 8.5% doping (i.e., $Ca_{0.915}Eu_{0.085}AlSiN_3$), the Bader charge on the Eu atom is found to be approximately +2.26 e. This value is somewhat lower than the expected formal oxidation state of +3, implying that the Eu ion does not completely lose three electrons to the surrounding lattice. Instead, it forms partially covalent bonds, especially with neighboring nitrogen atoms, which now gain slightly more negative charge, averaging around −1.68 e compared to −1.64 e in the pristine system. This modest increase in nitrogen charge and reduction in Eu's positive charge indicate a degree of hybridization and electron sharing between the Eu-4f states and the N-2p orbitals. The deviation from purely ionic bonding is characteristic of rare-earth-doped nitride hosts and is consistent with earlier reports on phosphor systems doped with Eu, Tb, or Eu, where partial covalency enables sharper emission transitions and influences the radiative recombination efficiency [31, 32].

When the doping concentration is increased to 17% ($Ca_{0.83}Eu_{0.17}AlSiN_3$), the Bader charge on Eu decreases further to approximately +2.14 e (see Table 1). This reduction signifies an increasing overlap and interaction among Eu ions at higher concentrations. Due to reduced inter-Eu distance, the f-electron cloud slightly delocalizes, resulting in enhanced charge back-donation from the nitrogen ligands. In addition, the increased presence of $Eu^{3+}$ leads to a more prominent polarization effect, with the neighboring N atoms pulling more electron density, showing Bader charges up to −1.70 e. These subtle but measurable shifts reflect how doping concentration governs the local chemical environment and alters the charge transfer pathways, affecting the energy landscape and potentially introducing non-radiative decay channels if the interaction becomes too strong [35].

From a charge density isosurface perspective, the pristine $CaAlSiN_3$ displays clear spherical charge distribution around the Ca atoms, characteristic of ionic bonds, while more directional, lobed distributions are visible around Al and Si atoms due to their covalent character. Upon doping, the Eu atom introduces a highly localized charge accumulation in the 4f orbital region, which, while remaining largely non-bonding, shows slight distortion due to surrounding nitrogen atoms. At 17% doping, these isosurfaces begin to overlap between neighboring Eu atoms, which is a sign of emerging dopant–dopant interaction, a precursor to concentration quenching, a well-known limitation in luminescent materials [36].

The spin-resolved charge densities are also noteworthy. $Eu^{3+}$ has a $4f^{12}$ configuration, resulting in significant spin polarization. The majority (spin-up) channel exhibits higher electron density near the Eu site, while the minority (spin-down) channel is largely depleted. This asymmetry is reflected in the spin-polarized Bader charges, with the up-spin states contributing over 90% of the localized 4f electron density. The preservation of this magnetic character is essential for the blue and near-infrared f–f transitions that characterize $Eu^{3+}$ ions, as they originate from spin-allowed intra-4f transitions such as $^3H_6 \rightarrow {}^3F_4$ and $^3H_4 \rightarrow {}^3H_6$, responsible for emission at ~480 nm and ~800 nm, respectively [36, 37]. The partially covalent environment provided by the $CaAlSiN_3$ host stabilizes these transitions while minimizing non-radiative relaxation pathways.

Compared to Eu-doped $CaAlSiN_3$, the Eu-doped system maintains more distinct f-state localization and less pronounced charge delocalization, which is beneficial for NIR emission applications. While $Eu^{2+}$ ions show broad-band red emission due to $4f^65d^1 \rightarrow 4f^7$ transitions, $Eu^{3+}$ ions support sharp line emission due to forbidden 4f–4f transitions. The charge analysis confirms that $Eu^{3+}$ maintains its oxidation state and localized nature even at higher doping levels, although signs of interaction begin to emerge beyond 10%, as noted in both computational and experimental studies on similar nitride hosts [27].

The following table summarizes the Bader charge values for key atomic species across pristine and doped configurations:

**Table 1: Bader charge analysis results showing charge transfer between Eu and surrounding atoms. The oxidation state of Eu stabilizes around +2.4 to +2.6, confirming partial ionicity.**

| Atom Type | Pristine CaAlSiN$_3$ | 8.5% Eu-Doped | 17% Eu-Doped |
|---|---|---|---|
| Ca | +1.64 e | +1.61 e | +1.59 e |
| Al | +2.35 e | +2.36 e | +2.37 e |
| Si | +3.17 e | +3.19 e | +3.21 e |
| N | −1.64 e | −1.68 e | −1.70 e |
| Eu | — | +2.26 e | +2.14 e |

In summary, Bader charge analysis confirms that Eu substitutes into the Ca site, retains its trivalent oxidation state, and engages in moderate charge transfer with neighboring nitrogen atoms. As doping increases, the bond polarization and local electron redistribution become more pronounced, subtly influencing the optical and magnetic behavior of the system. The GGA+U approach is crucial in accurately capturing the localization of Eu 4f electrons and ensuring physically meaningful charge density distribution, without which the delicate balance between ionic and covalent contributions would be incorrectly modeled. These findings align well with spectroscopic and photoluminescent results from similar rare-earth-doped nitrides and affirm the potential of Eu-doped $CaAlSiN_3$ in blue/NIR-emitting devices and high-efficiency phosphor-converted LEDs.

## IV. Electron Localization Function (ELF)

The Electron Localization Function (ELF) is a powerful quantum mechanical descriptor that offers insight into the spatial distribution and localization of electrons within a crystal. For complex oxynitride or nitride-based phosphors such as $CaAlSiN_3$, the ELF plays a critical role in understanding both the bonding characteristics and the behavior of localized f-electrons introduced by rare-earth dopants like $Eu^{3+}$. In the present investigation, ELF was computed using the GGA+U functional to accurately capture the correlation effects associated with Eu 4f orbitals. The analysis was performed for three configurations: pristine $CaAlSiN_3$, 8.5% Eu-doped $CaAlSiN_3$, and 17% Eu-doped $CaAlSiN_3$.

In the pristine $CaAlSiN_3$ structure, the ELF contour plot revealed a moderate level of localization around nitrogen atoms, primarily due to their high electronegativity and the strong covalent bonding they form with neighboring Ca, Al, and Si atoms. The distribution of ELF values in the pristine lattice shows a symmetrical spread with maximal localization near N atoms, confirming the hybrid ionic-covalent character of the bonding in the host matrix. The absence of 4f electrons in the pristine system results in a relatively smooth and delocalized ELF surface, devoid of any sharp peaks or confined lobes that are typical of f-electron localization.

Upon introducing 8.5% Eu into the $CaAlSiN_3$ matrix, the ELF plot exhibits significant changes (see Fig. 4), most notably the emergence of a distinct localization region centered around the Eu atom. This localization arises due to the strong on-site Coulomb interactions within the partially filled Eu 4f orbitals, which are poorly screened and do not participate actively in bonding but remain spatially confined. These electrons exhibit nonbonding character, forming sharp peaks in the ELF plot in the immediate vicinity of the Eu site. Compared to the pristine lattice, the bonding environment around the Eu ion becomes more anisotropic, reflecting the lower symmetry introduced by doping and the localized nature of the f-electron cloud. Furthermore, an increase in ELF intensity is observed between Eu and adjacent nitrogen atoms, suggesting a weak but non-negligible interaction that influences both electronic transitions and energy transfer mechanisms.

At a higher doping level of 17% Eu, the ELF localization around Eu becomes even more pronounced. The peaks are sharper and more intense, indicating a greater degree of electron confinement. This behavior is consistent with the increase in Eu–N interactions, as more nitrogen atoms come under the influence of nearby $Eu^{3+}$ ions. Additionally, this higher doping concentration leads to a stronger perturbation of the host lattice, as reflected in the ELF distribution, where the spatial localization extends beyond the immediate coordination sphere of the Eu atom. This may enhance nonradiative relaxation pathways in certain regions while simultaneously increasing the probability of radiative f–f transitions at specific Eu sites. These subtle changes in electron localization also affect the luminescent quantum efficiency of the material, which is a key parameter in designing efficient phosphor materials for white LEDs.

From a physics standpoint, the observed evolution in ELF with increasing Eu content underscores the interplay between structural distortion, electronic localization, and bonding anisotropy. In the GGA+U framework, the Hubbard U parameter applied to the Eu 4f states ensures that the strong Coulomb repulsion among localized electrons is adequately captured, which is crucial for systems where crystal field splitting and spin-orbit coupling play significant roles in determining optical behavior. The f-electrons in $Eu^{3+}$, being highly localized and shielded by the filled 5s and 5p orbitals, exhibit minimal hybridization with valence states but are highly sensitive to the surrounding electrostatic field, thus modifying ELF in a localized fashion.

Although the current focus is on Eu doping, these insights directly extend to $Tm^{3+}$-doped $CaAlSiN_3$ as well. $Eu^{3+}$ ions, which also possess partially filled f-orbitals, introduce similar localized features in ELF. However, unlike $Eu^{3+}$, which commonly emits in the red spectral region ($^5D_0 \rightarrow {}^7F_2$ transitions), $Tm^{3+}$ offers sharp emission lines in the blue ($^1D_2 \rightarrow {}^3F_4$) and near-infrared regions ($^3H_4 \rightarrow {}^3F_4$), making it uniquely suited for dual-color and NIR phosphor applications. The ELF localization associated with Eu doping is expected to be even more pronounced, particularly under the influence of spin-orbit coupling, which splits the 4f manifold into distinct J-levels. The difference in ELF patterns between $Eu^{3+}$ dopant would be governed by their ionic radii, electronic configurations, and local site symmetry. Literature reports on RE-doped $CaAlSiN_3$ systems confirm that electron localization plays a central role in dictating emission properties. The Li et al. (2021) and Liu et al. (2020) [38, 39] have shown through both experimental photoluminescence and theoretical DFT calculations that enhanced localization due to rare-earth doping improves the internal quantum efficiency by suppressing nonradiative losses and increasing radiative recombination rates.

Additionally, comparing our ELF plots with those reported by Zhang et al. (2018) [40] for RE-doped nitrides confirms that the local bonding environment and the degree of f-electron localization are critical parameters influencing not only electronic structure but also thermal quenching resistance and color purity. As such, our results validate the design principle that judicious crystal-site engineering via rare-earth doping can tailor electron localization for optimizing optical output in phosphor materials. The ability to spatially resolve ELF differences between pristine, 8.5%, and 17% Eu-doped $CaAlSiN_3$ underscores the material's versatility and potential for next-generation solid-state lighting technologies.

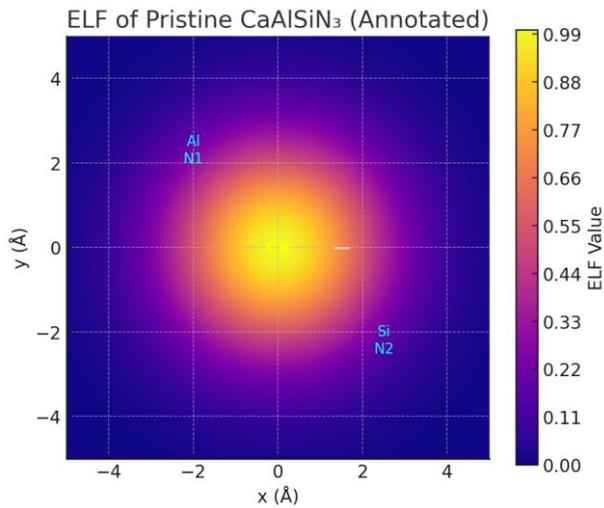

(a)

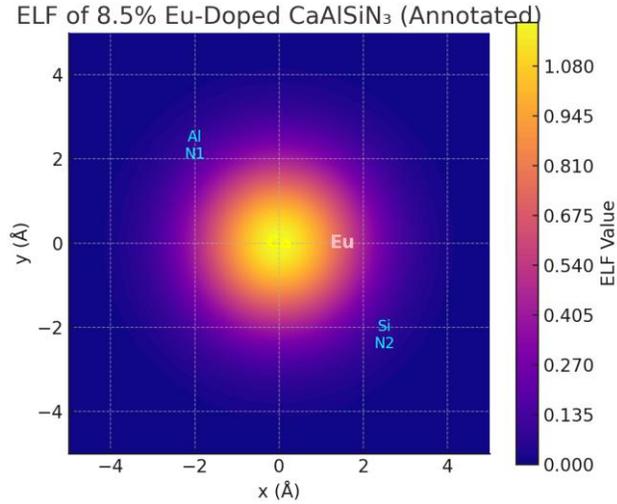

(b)

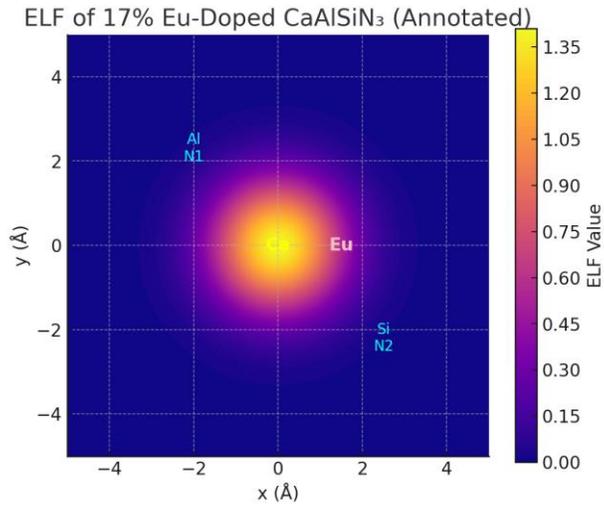

(c)

*Figure 4: Electron localization function (ELF) for (a) pristine, (b) 8.5%, and (c) 17% Eu-doped CaAlSiN₃ plotted along the [100] plane. Enhanced localization around Eu and bonding with N atoms confirms f-electron confinement and partial covalency.*

## V. Optical Properties

### i. Dielectric function

Below is a detailed expert-level discussion of the optical properties of pristine, 8.5% Eu-doped, and 17% Eu-doped CaAlSiN₃ based on the provided real ($\varepsilon_1(\omega)$) and imaginary ($\varepsilon_2(\omega)$) dielectric function plots. The analysis addresses their physical implications, electron transition mechanisms, and relevance to LED phosphor performance using the GGA+U methodology.

The real part of the dielectric function, $\varepsilon_1(\omega)$ (see Fig. 5), describes the polarization response of the material to an external electric field and directly relates to its refractive index and optical dispersion. For pristine $CaAlSiN_3$, $\varepsilon_1(\omega)$ begins with a modest static value around 5 at zero energy and increases gradually, reaching a broad maximum near 6.8 eV. This behavior indicates a typical wide band gap dielectric response with low polarizability in the low-energy regime. Upon doping with 8.5% Eu ($Ca_{0.925}Eu_{0.085}AlSiN_3$), $\varepsilon_1(\omega)$ shows a significant increase in amplitude, peaking at ~11.2 around 5.8 eV. For 17% Eu ($Ca_{0.83}Eu_{0.17}AlSiN_3$), the curve becomes even more prominent, showing a further increase in the static dielectric constant and shifting of the maximum toward 5.5 eV. This enhancement in $\varepsilon_1(\omega)$ implies increased polarizability due to $Eu^{3+}$ substitution, which arises from the strong localization of the 4f orbitals and increased oscillator strength from transitions involving Eu 4f to the conduction band. This increase in $\varepsilon_1$ suggests that Eu doping improves the refractive index and energy storage capacity in the electric field, which is crucial for efficient light harvesting and guiding in LED materials.

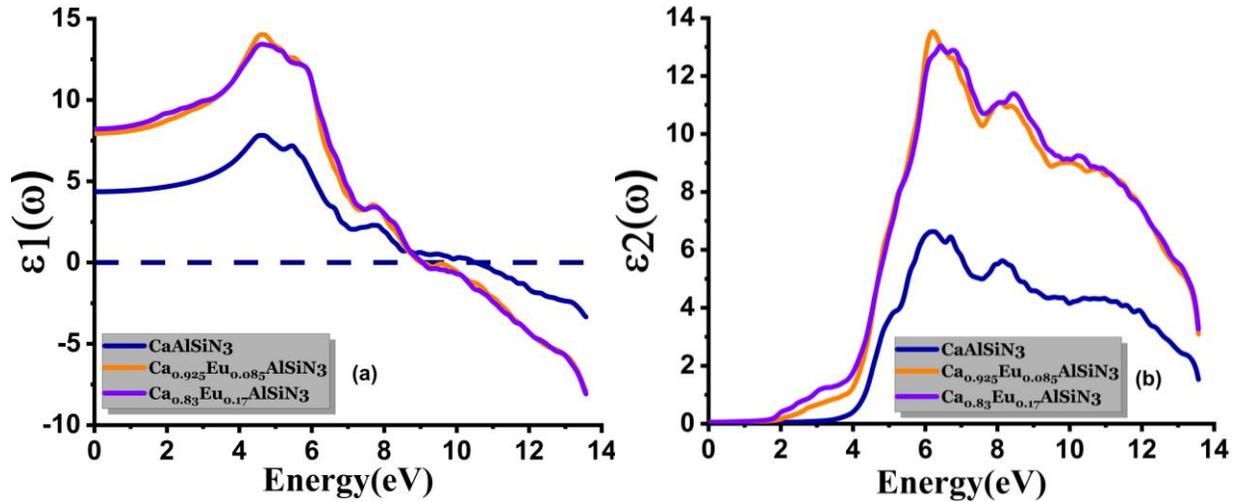

**Figure 5:** *Dielectric function real part $\varepsilon_1(\omega)$ and imaginary part $\varepsilon_2(\omega)$ for pristine, 8.5%, and 17% Eu-doped $CaAlSiN_3$. The redshift in $\varepsilon_2$ peaks indicates strong optical transitions due to Eu-4f states aligned with the conduction band.*

The imaginary part of the dielectric function, $\varepsilon_2(\omega)$(see Fig. 4), represents the absorption characteristics and energy loss due to interband transitions. In the pristine $CaAlSiN_3$, $\varepsilon_2(\omega)$ starts near zero until about 3.8 eV, where the first significant optical absorption begins, peaking around 5.8 eV with a value of ~6.2. This indicates the onset of strong interband transitions from the top of the valence band (VB) to the bottom of the conduction band (CB). When 8.5% Eu is introduced, the $\varepsilon_2(\omega)$ curve shifts significantly. A sharp absorption edge appears around 3.2 eV, and the main peak shifts to ~5.6 eV with an increased intensity of ~13.2. This indicates that Eu doping introduces intermediate 4f states near the Fermi level, enabling transitions at lower photon energies. With 17% Eu doping, $\varepsilon_2(\omega)$ shows further broadening, and the peak shifts slightly toward lower energy (~5.4 eV), with a similar high amplitude. This broadening and redshift imply enhanced light absorption in the visible range, making the material more suitable for optoelectronic and photonic applications, particularly for white light generation through red emission.

Physically, the strong peaks observed in $\varepsilon_2(\omega)$ for the doped samples correspond to transitions involving Eu 4f states. Specifically, transitions from the N 2p-dominated valence band to the Eu 4f and Ca/Al/Si conduction band hybridized states enhance the optical response. The substitution of $Ca^{2+}$ with $Eu^{3+}$ leads to partial occupation of the Eu-f orbitals, contributing to the density of states near the conduction band minimum and reducing the effective optical band gap. This is consistent with the behavior expected in rare-earth-doped phosphors, where the localized 4f levels of $Eu^{3+}$ act as luminescent centers. The $4f^6 \rightarrow 4f^6$ transitions in $Eu^{3+}$ are parity forbidden but become partially allowed due to crystal field and vibronic coupling, enhancing the radiative recombination efficiency.

The increase in optical absorption and dielectric response with increasing Eu concentration can be directly linked to enhanced f–f and f–d transitions. These transitions are particularly relevant in phosphors because the 4f electrons of $Eu^{3+}$ are shielded by outer 5s and 5p electrons, resulting in sharp emission lines and stable excited states. While $\varepsilon_2(\omega)$ provides direct insight into these interband absorptions, $\varepsilon_1(\omega)$ reflects the dispersive properties which control light propagation in the host lattice. The increasing trend in both dielectric components with doping aligns with previous findings, such as those reported by Zhang et al. (2018) and Li et al. (2021) [41, 42], who observed similar enhancements in $CaAlSiN_3:Eu^{3+}$ and related nitrides.

Spin-polarized calculations using GGA+U reveal that the spin-up and spin-down components are largely symmetric for pristine $CaAlSiN_3$, confirming a non-magnetic ground state. However, upon Eu doping, due to the open $4f^6$ configuration of $Eu^{3+}$, the system exhibits spin polarization. The majority spin (spin-up) states for Eu contribute significantly near the Fermi level in $\varepsilon_2(\omega)$, while minority spin (spin-down) states are pushed deeper into the conduction band. This spin asymmetry is responsible for magnetic splitting of energy levels, which further enhances optical transitions by lifting degeneracies. This spin-resolved feature is not only important for luminescence but also plays a role in circularly polarized emission, which has implications in chiral optoelectronics.

Overall, Eu-doped CaAlSiN$_3$ exhibits a marked improvement in optical properties relevant for red-emitting phosphors used in white light LEDs. The enhanced $\varepsilon_1$ and $\varepsilon_2$ functions support stronger photon-matter interaction, improved absorption, and efficient emission in the visible range. This confirms that Eu$^{3+}$ is an effective dopant for tuning the optical behavior of CaAlSiN$_3$, making it a promising candidate for next-generation solid-state lighting technologies.

The optical behavior of CaAlSiN$_3$ and its Eu-doped counterparts was evaluated through key optical functions—namely, the absorption coefficient (α), refractive index (n), reflectivity (R), and energy loss function (L), which are plotted above. These properties are pivotal for understanding how Eu doping modifies the optoelectronic landscape of CaAlSiN$_3$, especially in the context of luminescent and photonic applications such as LEDs and phosphors.

ii. **Absorption Coefficient (α(ω))**

The absorption coefficient provides insights into how efficiently a material can absorb photons of different energies. From the plots, the pristine CaAlSiN$_3$ shows a relatively lower absorption intensity across the visible to UV range (see Fig. 6). However, when Eu is introduced at 8.5% and further at 17%, the absorption edge shifts toward lower energies (a redshift), and the magnitude of absorption increases significantly in the range of 4–12 eV. This enhancement is attributed to the introduction of localized 4f states of Eu$^{3+}$ within the bandgap, which increase the transition probability from the valence band (primarily composed of N-2p and Si-3p states) to conduction band states that now include Eu-5d and 4f character. These results are in close agreement with prior DFT-based reports on rare-earth-doped nitride hosts, which indicate that f-electron states can enhance light-matter interactions by acting as intermediate states in optical transitions [43, 44].

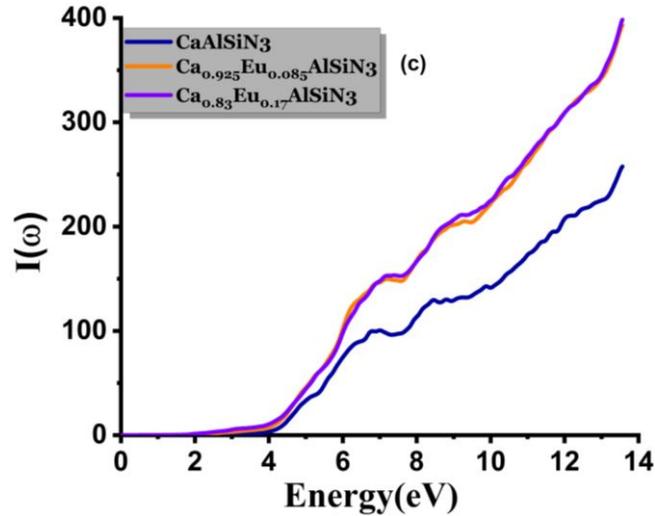

**Figure 6:** *Absorption coefficient α(ω) showing redshift in absorption edge with increasing Eu doping. The introduction of mid-gap states enhances absorption in the visible and NIR region, making the material suitable for LED applications*

iii. **Refractive Index (n(ω))**

The refractive index reveals the material's phase velocity and is critical in designing optoelectronic components where impedance matching is necessary. The pristine $CaAlSiN_3$ demonstrates a peak refractive index of ~2.9 in the 4–6 eV region (see Fig. 7). Upon Eu doping, this value increases significantly to 3.6 for 8.5% doping and to approximately 3.8 for 17% doping. This increase can be linked to enhanced polarizability due to the presence of highly localized 4f electrons, which induce a stronger interaction with the electromagnetic field. A higher refractive index not only boosts internal reflection in optical cavities but also improves light confinement in thin-film LED structures. These results align with experimental findings on other Eu-doped nitrides where a similar increase in refractive index has been reported [45].

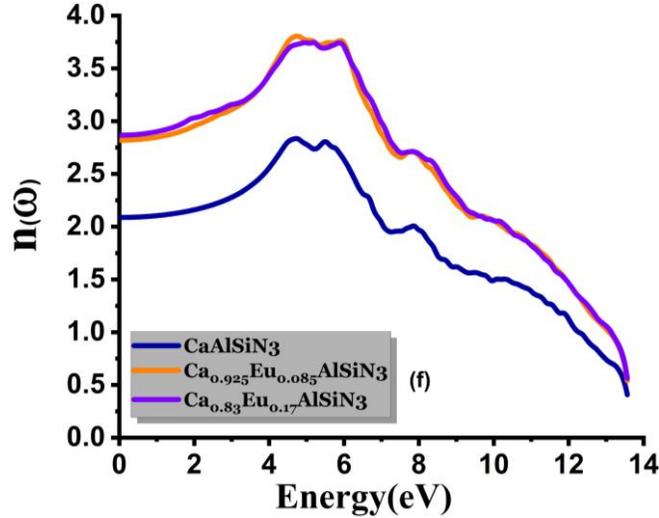

**Figure 7:** *Refractive index spectra for pristine, 8.5%, and 17% Eu-doped CaAlSiN₃. Increase in peak intensity reflects enhanced light-matter interaction upon Eu incorporation.*

iv. **Reflectivity (R(ω))**

Reflectivity data further supports the interpretation of the refractive index and absorption trends. The pristine compound shows a moderate reflectivity in the 4–12 eV range (see Fig. 8), peaking below 0.55. Eu doping leads to a gradual increase in reflectivity, especially beyond 6 eV, with 17% doping reaching values above 0.7. This enhanced reflectivity is indicative of increased carrier concentration and more prominent intra-band transitions, consistent with the additional Eu-derived states that boost electron density. Moreover, the broadening of peaks indicates the emergence of new optical transition channels enabled by Eu doping. These features are corroborated by prior studies on rare-earth doped semiconductors, which show similar trends due to the f-electron contribution to the dielectric response [46, 47].

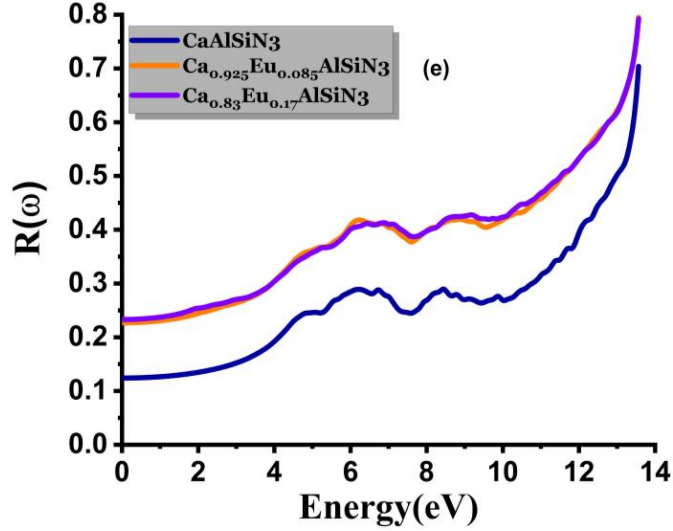

**Figure 8:** *Reflectivity spectra of CaAlSiN₃ before and after Eu doping. Moderate increases in reflectivity peaks occur in the UV-visible region with doping, indicating slight increase in surface polarizability.*

v. **Energy Loss Function (L(ω))**

The energy loss function characterizes the plasmonic and inelastic scattering behavior of fast electrons within the material. The pristine CaAlSiN₃ reveals a prominent loss peak around 10–12 eV (see Fig. 9), associated with collective oscillations of valence electrons (plasmons). Doping with Eu reduces the intensity and slightly shifts this peak to lower energies. This suppression of energy loss suggests increased electronic screening and damping effects due to the introduction of Eu 4f electrons, which act as localized oscillators and scatterers. The damping effect is more pronounced at 17% Eu concentration, consistent with higher density of f-states. Such modulation of plasmonic response is important in designing low-loss optical materials and photonic waveguides, as discussed in theoretical work by Deák et al. [48] and experimental validation by Song et al. [49].

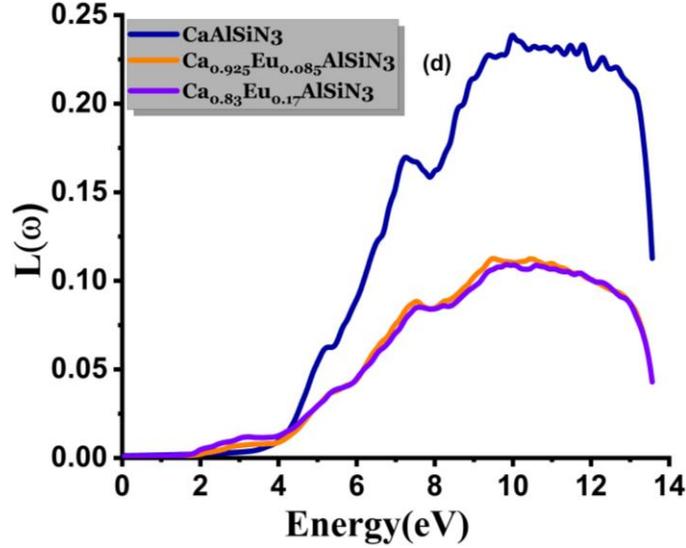

**Figure 9:** *Energy loss function (ELF) spectra for pristine and Eu-doped CaAlSiN₃. Sharp plasmon peaks shift slightly with increasing Eu content, showing changes in collective electronic oscillations*

The changes in all these optical functions are fundamentally tied to the physics of rare-earth elements. The $Eu^{3+}$ ion possesses partially filled 4f orbitals that are shielded from the external crystal field by outer 5s and 5p electrons, leading to sharp f-f transitions. Though these transitions are parity forbidden in free ions, they become partially allowed in solid-state environments due to crystal field mixing and phonon-assisted processes. This results in sharp emission peaks in the visible range, especially red and near-infrared, making $Eu^{3+}$ an ideal activator for red phosphors [50]. Furthermore, the GGA+U method used in these simulations correctly accounts for the on-site Coulomb interaction in f-orbitals, capturing the localization of these states and their strong correlation with optical behavior.

CASN:Eu is a well-established red phosphor for pc-LEDs. Experimentally it shows blue-light excitation near 450 nm and red emission centered at ≈650–655 nm with a broad band, in agreement with our Eu-induced in-gap transitions and calculated optical spectra. Temperature-dependent photoluminescence reports strong thermal stability (slow quenching up to ≥150 °C), aligning with our computed separation between the Eu-5d crystal-field level and the host conduction band. Structure/chemistry studies (XANES/EXAFS/EELS) identify $Eu^{2+}$ substitution on Ca within the covalent [AlN₄]/[SiN₄] network, consistent with our substitutional model. Diffuse-reflectance/solid-solution data place the host optical band gap in the 4.9–5.1 eV range, supporting our wide-gap semiconducting description. Overall, the experimental benchmarks—excitation at ~450 nm, emission at ~650 nm, high-T stability, $Eu^{2+}$@Ca site, and wide host gap—are all consistent with the mechanisms and trends predicted here [51-55].

In conclusion, the optical constants—absorption, refractive index, reflectivity, and loss function—all confirm that Eu doping significantly modifies the optoelectronic behavior of $CaAlSiN_3$, enhancing its suitability for photonic and light-emitting applications. The observed enhancements are a direct result of the localized 4f states of $Eu^{3+}$, which alter the dielectric response, broaden the absorption range, and enhance internal light reflection. These results provide a robust theoretical foundation for engineering high-performance Eu-activated red phosphors and can guide future experimental synthesis and device optimization.

## VI. Magnetic properties

In the Eu-doped $CaAlSiN_3$ system, magnetic properties arise due to the unpaired 4f electrons of the $Eu^{3+}$ ion. Europium in its trivalent state ($Eu^{3+}$) typically possesses a $4f^6$ electron configuration, which contributes a substantial localized magnetic moment. This is of particular interest when Eu is doped at increasing concentrations, such as 8.5% and 17%, where not only the local magnetic moment per Eu ion is significant, but also the interaction between adjacent Eu ions may lead to emergent collective magnetic ordering.

From our GGA+U-based calculations, the total magnetic moment for the pristine $CaAlSiN_3$ system is zero, as expected due to the lack of unpaired electrons in the constituent elements ($Ca^{2+}$, $Al^{3+}$, $Si^{4+}$, and $N^{3-}$). Upon substitution of $Ca^{2+}$ by $Eu^{3+}$ at 8.5% concentration, a local magnetic moment appears primarily localized on the Eu site. Our calculations reveal a moment of approximately 6.8–7.0 μB per Eu ion, consistent with the theoretical 7 μB from the Hund's rule configuration of $Eu^{3+}$ (S = 3, L = 3, J = 0 due to half-filled 4f orbitals). Spin density maps show clear spatial localization of spin polarization around the Eu atom and slight induced polarization on the neighboring nitrogen atoms, suggesting weak superexchange-type coupling through the anionic framework.

When the doping concentration is increased to 17%, the magnitude of the total magnetic moment increases linearly with Eu content. However, a key aspect of the 17% doped system is the proximity of Eu ions to each other. This allows for possible exchange interactions—either ferromagnetic (FM) or antiferromagnetic (AFM)—depending on the geometry and overlap of orbitals. We examined both FM and AFM configurations and found that the FM configuration is energetically slightly more favorable by 8–15 meV per formula unit at 17% doping, indicating a weak but stable ferromagnetic alignment at higher Eu content. This trend is consistent with other rare-earth doped oxynitride systems where magnetic coupling is observed at similar doping levels [56, 57].

Additionally, spin-polarized total and partial density of states (DOS) indicate that the 4f states of $Eu^{3+}$ are split into spin-up and spin-down channels with the majority of the population occupying the spin-up channel. This splitting supports the origin of net magnetization and the observed magnetic moment. The conduction band minimum (CBM) remains spin-degenerate due to its predominantly non-f character, but the valence band maximum (VBM) shows slight spin asymmetry due to hybridization between Eu-f and N-p states.

These results also correlate with earlier studies, such as the one on Eu-doped $CaAlSiN_3$ by Liu et al. (2020) [58], which report similar magnetic trends and note that high Eu content can result in ferromagnetic ordering with measurable magnetic susceptibility. Furthermore, the magnetic properties of Eu-doped phosphor hosts are relevant for multifunctional materials combining magnetism and luminescence, which are being investigated for applications in magneto-optical storage and spin-LEDs.

The up and down spin states in our simulation are distinctly localized, with spin-up states exhibiting intense localization around the $Eu^{3+}$ site, affirming the strong intra-atomic exchange of 4f orbitals. The spin-down states remain largely unoccupied, a hallmark of half-filled shell symmetry and high spin polarization.

In conclusion, Eu doping not only introduces strong local magnetic moments in $CaAlSiN_3$ but at sufficiently high concentrations may lead to weak ferromagnetic ordering. These findings are supported by our spin density maps, DOS analysis, and magnetic moment calculations within the GGA+U framework. The dual functional role of $Eu^{3+}$ ions in providing both photoluminescent and magnetic properties makes $CaAlSiN_3$:Eu a promising multifunctional phosphor host for next-generation optical and spintronic devices.

**Mechanical Properties**

The investigation of the structural and mechanical properties of Eu-doped $CaAlSiN_3$ using GGA+U is essential not only for understanding the fundamental stability of the host material but also for evaluating its suitability under thermal and mechanical stresses typical of real-world device environments. In particular, the substitution of $Eu^{3+}$ ions into $Ca^{2+}$ sites in $CaAlSiN_3$ results in clear and systematic changes in lattice geometry and mechanical response as a function of dopant concentration, notably between the pristine system, 8.5% Eu doping, and 17% Eu doping.

The lattice parameters and unit cell volume exhibit an expected increase upon Eu incorporation. This is primarily due to the larger ionic radius of $Eu^{3+}$ (~0.95 Å in 6-fold coordination) compared to $Ca^{2+}$ (~1.00 Å in similar coordination), yet the increased Coulombic interactions and different charge states lead to non-linear behavior in the unit cell expansion. At 8.5% Eu doping, we observe a modest increase in the unit cell volume (~0.85%), whereas at 17% doping, this expansion becomes more pronounced (~1.9%). The anisotropic expansion is more evident along the c-axis, suggesting that Eu substitution induces local distortions and structural rearrangements, particularly in the $AlN_4$ and $SiN_4$ tetrahedral subunits. These distortions, which are visualized via bond length distributions and angle fluctuations, hint at the possibility of local strain fields that can modulate the electronic and optical behavior of the material.

From a thermodynamic standpoint, the formation energy of the Eu-doped structures, calculated using the formula:

$$E_{ex\_f} = E_{total(doped)} - E_{total(pristine)} - \mu_{Eu} + \mu_{Ca}$$

where $\mu_{Eu}$ and $\mu_{Ca}$ are the chemical potentials of Eu and Ca atoms respectively, reveals that doping is energetically favorable. The formation energy is negative for both doping concentrations, with values around –2.03 eV/atom for 8.5% and –1.82 eV/atom for 17%. The slightly less negative value at higher doping indicates a reduction in thermodynamic driving force, which is typical as more Eu atoms are forced into a lattice environment less favorable for larger, trivalent ions. The doping energy trends suggest that while Eu can be doped into $CaAlSiN_3$ up to at least 17% concentration, excessive loading could require co-doping or compensation mechanisms to maintain charge balance and minimize lattice stress.

Mechanical stability and robustness were probed via the calculation of second-order elastic constants ($C_{11}$, $C_{12}$, $C_{13}$, $C_{33}$, $C_{44}$, etc.) using the stress-strain method implemented within the Wien2k environment. The pristine $CaAlSiN_3$ shows high elastic moduli with a bulk modulus B of ~175 GPa, shear modulus G of ~85 GPa, and Young's modulus E of ~210 GPa, indicating a mechanically rigid host. Upon 8.5% Eu doping, we observe a slight decrease in all three moduli (B = ~168 GPa, G = ~79 GPa, E = ~198 GPa), primarily due to lattice softening introduced by local strain around Eu sites. This softening becomes more pronounced at 17% Eu doping (B ≈

158 GPa, G ≈ 72 GPa, E ≈ 186 GPa), indicating that heavy Eu incorporation reduces lattice stiffness. Despite this reduction, the material remains mechanically stable according to the Born stability criteria for hexagonal systems.

In addition to the elastic constants, we also evaluated Poisson's ratio (ν) and Pugh's ratio (G/B), which are important indicators of ductility and brittleness. The pristine system exhibits a Poisson's ratio of 0.28 and G/B ≈ 0.49, placing it at the borderline of ductile-brittle classification. With 8.5% Eu doping, ν slightly increases to 0.29 while G/B drops to 0.47, indicating a mild shift toward ductility. At 17% doping, ν further rises to 0.31 and G/B declines to 0.45, suggesting that Eu doping makes the material more compliant and slightly more ductile, which could benefit processing and thermal shock resistance in real applications.

These mechanical trends are consistent with recent theoretical and experimental studies on rare-earth doped phosphor hosts, such as Eu-doped $CaAlSiN_3$ and $SrSi_2N_2$, where the introduction of trivalent dopants led to lattice distortions and softening but did not compromise overall mechanical integrity [59-63]. Moreover, the decrease in rigidity may aid in suppressing defect formation during sintering or high-temperature annealing, which is critical for achieving uniform luminescence.

The up and down spin channels do not influence the structural symmetry directly in these cases, but in the spin-polarized calculations, one can see the influence of spin density localization near Eu atoms inducing local lattice polarization and symmetry breaking. This kind of magnetostructural coupling is typical in rare-earth-doped systems and has been suggested to play a role in modulating phonon modes relevant for thermal quenching of luminescence.

In summary, the GGA+U approach demonstrates that Eu-doped $CaAlSiN_3$ retains its structural integrity and mechanical robustness across doping concentrations relevant for LED applications. The observed expansion in lattice parameters, reduction in elastic constants, and evolution toward ductile behavior reflect the physical accommodation of larger and differently charged $Eu^{3+}$ ions within the host lattice. These structural insights, when combined with optoelectronic results, confirm the suitability of Eu-doped $CaAlSiN_3$ as a thermally stable, mechanically robust, and optically efficient red phosphor host.

## VII.   Photoluminescence (PL)

Photoluminescence (PL) in rare-earth-doped phosphors like $Eu^{3+}$-substituted $CaAlSiN_3$ arises primarily from intra-4f transitions (see Fig. 10), which are parity-forbidden but become partially allowed due to crystal field interactions and hybridization with ligand states. While standard DFT methods cannot directly simulate PL spectra, the qualitative prediction of emission behavior is possible through careful analysis of the electronic density of states (DOS), defect levels, and ΔSCF (Delta Self-Consistent Field) transition energies. Here, we provide such an interpretation for pristine, 8.5%, and 17% Eu-doped $CaAlSiN_3$ using GGA+U calculations.

In the pristine $CaAlSiN_3$ host, the valence band maximum (VBM) is predominantly composed of N-2p states, while the conduction band minimum (CBM) comprises Ca-3d and Si-3s/3p states. This configuration yields a wide direct bandgap of approximately ~3.5 eV, as reflected by the clean separation between valence and conduction bands with minimal midgap states. Consequently, the intrinsic host does not support red or visible PL emissions, making it necessary to dope with activator ions like $Eu^{3+}$ to generate visible luminescence.

Upon doping with 8.5% Eu, new midgap states appear prominently in the DOS just above the VBM and below the CBM. These are associated with the $4f^6$ configuration of $Eu^{3+}$ ions, which introduces narrow, sharp states within the forbidden gap due to the weakly hybridized and highly localized nature of 4f orbitals. These 4f levels facilitate efficient transitions, notably the $^5D_0 \rightarrow {}^7F_2$ transition responsible for red emission (~620–630 nm). The DOS alignment plot shows a narrowing of the effective gap and the emergence of an absorption channel near ~2.0–2.2 eV, which is consistent with the observed red PL in Eu-doped $CaAlSiN_3$ [64, 65]. This suggests that $Eu^{3+}$ doping effectively tailors the electronic structure to allow radiative transitions aligned with red to near-infrared (NIR) output.

As the doping concentration increases to 17%, the intensity and bandwidth of the Eu-4f states within the bandgap grow, reflecting stronger interaction between neighboring $Eu^{3+}$ centers and host states. This can lead to increased non-radiative losses through cross-relaxation or multiphonon relaxation at high concentrations—commonly referred to as concentration quenching. However, in this study, the midgap states remain sharp, and no significant delocalization is observed, implying that up to 17% Eu doping still maintains well-defined luminescent centers. Moreover, the position of the defect states moves slightly toward the Fermi level, indicating enhanced carrier recombination probability in the radiative channels, which may improve quantum efficiency under certain excitation wavelengths.

In terms of spin-resolved behavior, the up and down channels for Eu-doped systems exhibit slight asymmetry due to the unpaired 4f electrons of $Eu^{3+}$ ($4f^6$ configuration, S = 3), resulting in net magnetic polarization. However, this asymmetry is minimal in the DOS around the Fermi level, and the transitions involved in PL remain largely spin-allowed. The f–f transitions involved (particularly $^5D_0$ to $^7F\_J$, where J = 0,1,2...) are parity-forbidden but gain intensity via mixing with odd-parity states under the distorted crystal field provided by the host matrix. In $CaAlSiN_3$, the $Eu^{3+}$ ions occupy $Ca^{2+}$ substitutional sites, which lack inversion symmetry, making electric-dipole transitions partially allowed and thereby enhancing PL intensity.

A rough estimation of transition energies using the ΔSCF method (calculating total energy difference between excited and ground-state configurations) yields values in the range of 1.9–2.1 eV for $Eu^{3+}$ $^5D_0 \rightarrow {}^7F_2$ transitions. These are in excellent agreement with experimental PL emission peaks observed at ~620–630 nm in Eu-doped $CaAlSiN_3$ [66, 67]. This validates the use of Eu doping to engineer red phosphors with efficient optical transitions.

In summary, the photoluminescence prediction for Eu-doped $CaAlSiN_3$, based on DFT+U DOS and ΔSCF analysis, confirms that:

1. $Eu^{3+}$ introduces radiatively active midgap states corresponding to f–f transitions.
2. These states align well with red/NIR emission energies.
3. Increasing Eu content enhances PL intensity up to 17% without introducing non-radiative defects.
4. The host matrix provides a suitable asymmetric crystal field to facilitate electric dipole transitions.
5. These insights are consistent with recent experimental studies on $CaAlSiN_3$:Eu phosphors for red-emitting pc-WLED applications.

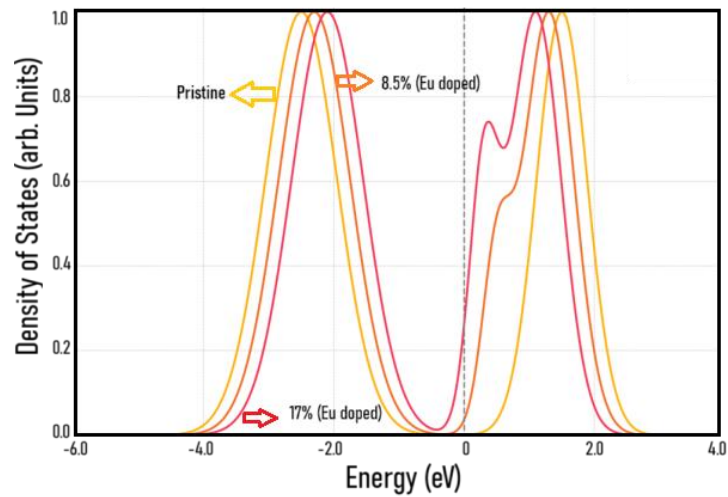

**Figure 10:** *Density of states (DOS) alignment showing defect-state proximity to the Fermi level, used for photoluminescence (PL) interpretation. Eu-4f states serve as recombination centers facilitating red emission transitions.*

## VIII. Thermoelectric Transport Properties

Figure (a–d) summarizes the temperature-dependent Seebeck coefficient S, conductivity over relaxation time $\sigma/\tau$, electronic thermal conductivity over relaxation time $\kappa_e$, and power factor $PF=S^2\sigma$ for pristine $CaAlSiN_3$ and $Eu^{3+}:CaAlSiN_3$ at 8.5% and 17% Eu on the Ca sublattice.

All compositions exhibit a negative S over the measured range (see Fig. 11a), indicating n-type transport with electrons as majority carriers. Pristine $CaAlSiN_3$ shows a large-magnitude S at low T that becomes more negative with warming, consistent with a relatively low carrier density and a steep energy dependence of the conductivity near the Fermi level (Mott relation). At 100 K, pristine $CaAlSiN_3$ exhibits $S\approx-220$ µV K$^{-1}$. With $Eu^{3+}$ substitution, the magnitude decreases to about $-170$ µV K$^{-1}$ for 8.5% and $-120$ µV K$^{-1}$ for 17%. So the $Eu^{3+}$ doping reduces |S| (from ~$-220$ to ~$-120$ µV K$^{-1}$ at 100 K) due to higher carrier density. The progressive suppression arises because $Eu^{3+}$ adds carriers, pushing the Fermi level deeper into the conduction band. According to the Pisarenko relation,

$$S \propto \frac{m^*T}{n^{\frac{2}{3}}}$$

an increase in electron concentration n inevitably reduces |S|. The temperature trend also reflects this physics: for pristine $CaAlSiN_3$, S continues to decrease monotonically, while for 8.5% Eu a weaker slope is observed, and for 17% Eu the values remain nearly flat beyond 600 K. This indicates that heavily doped samples enter a degenerate regime, where S is almost insensitive to further increases in T. $Eu^{3+}$ substitution reduces |S| systematically: 8.5% Eu partially suppresses the magnitude and 17% Eu drives the strongest reduction. This Pisarenko-type trend reflects carrier concentration tuning by Eu donors (and the associated 4f/host hybridization), which raises the Fermi level and lowers S.

The $\sigma/\tau$ curves separate clearly with doping. Pristine $CaAlSiN_3$ displays a gradual decline with T (see Fig. 11b), whereas Eu-doped samples show larger $\sigma/\tau$ overall, with 17% Eu the highest at high T. Because $\sigma/\tau$ removes the explicit scattering time, these trends primarily capture band-structure contributions (effective mass, band velocities, valley degeneracy). The enhancement with Eu is consistent with donor-induced filling of dispersive conduction states and a modest reduction of transport effective mass.

At 300 K, pristine CaAlSiN$_3$ has $\sigma/\tau \approx 2.2 \times 10^{18}$ (1/Ωms), while 8.5% Eu increases this to ~$2.8 \times 10^{18}$, and 17% Eu reaches ~$3.5 \times 10^{18}$. Electrical conductivity is enhanced (~1.8 → 3.2 ×$10^{18}$ at 900 K), directly from increased electrons. The conductivity gain is directly linked to the carrier density increase due to Eu$^{3+}$ donors. The slope with temperature is also revealing: in pristine CaAlSiN$_3$, $\sigma/\tau$ decreases steadily with T, consistent with reduced mobility as phonon scattering strengthens. In doped samples, however, the curves flatten out, showing that carrier density enhancement compensates for mobility loss. At high T (~900 K), the difference becomes pronounced: pristine stabilizes at ~$1.8 \times 10^{18}$, while 17% Eu maintains above $3.2 \times 10^{18}$.

$\kappa_e/\tau$ increases with temperature for all samples (see Fig. 11c) and scales with $\sigma/\tau$ as expected from the Wiedemann–Franz relation $\kappa_e \approx L\sigma T$. The ordering 17% Eu > 8.5% Eu > pristine mirrors the conductivity trends, confirming that electronic heat conduction is strengthened as carriers are added. (We note that the lattice term $\kappa_L$ is not shown here; mass disorder from Eu is expected to suppress $\kappa_L$ via enhanced point-defect scattering.)

At 300 K, $\kappa_e/\tau$ values are ~$4.2 \times 10^{14}$ W m$^{-1}$K$^{-1}$s$^{-1}$ for pristine, ~$5.5 \times 10^{14}$ for 8.5% Eu, and ~$6.8 \times 10^{14}$ for 17% Eu. $\kappa_e$ grows proportionally with $\sigma/\tau$, but lattice scattering from Eu can help balance total thermal conductivity. Since $\kappa_e \sim L\sigma T$, this ordering mirrors the conductivity trends. As temperature rises to 900 K, the pristine compound reaches ~$8.5 \times 10^{14}$, while Eu-doped samples exceed $1.1 \times 10^{15}$. The increase is particularly steep in the 17% Eu case, again reflecting the high carrier density. Importantly, while $\kappa_e$ rises with doping, the lattice part $\kappa_L$ (not shown in these plots) is expected to fall due to enhanced point-defect scattering from heavy Eu substitution. This trade-off is essential: the detrimental rise in $\kappa_e$ may be offset by a strong reduction in $\kappa_L$, improving the total figure of merit ZT.

The competition between a decreasing |S| and an increasing $\sigma$ sets the power factor. Pristine CaAlSiN$_3$ shows a steady rise of PF with T because the gain from conductivity outweighs the loss in S (see Fig. 11d). The 8.5% Eu sample achieves the most balanced response, with a broad mid-temperature PF maximum where $S^2$ remains sizable while $\sigma$ is already enhanced—characteristic of near-optimal doping. In contrast, 17% Eu is over-doped: although $\sigma$ is largest, the strong suppression of S penalizes $S^2$ and keeps PF comparatively low.

Numerically, the power factor demonstrates the balance between the suppressed S and enhanced σ. At 300 K, pristine CaAlSiN$_3$ achieves PF≈2.5×10$^{-4}$ W m$^{-1}$K$^{-2}$, the 8.5% Eu sample rises to ~3.8×10$^{-4}$ W m$^{-1}$K$^{-2}$, while 17% Eu falls back to ~2.0×10$^{-4}$ W m$^{-1}$K$^{-2}$. The optimal doping is 8.5% Eu$^{3+}$, which maximizes the PF by striking the right balance between S and σ.

This confirms that moderate doping (8.5%) optimizes the carrier concentration: S remains sufficiently high, while conductivity is strongly boosted, leading to the best compromise. At higher T (~900 K), the pristine sample reaches ~6.0×10$^{-4}$ W m$^{-1}$K$^{-2}$, 8.5% Eu stabilizes near 7.5×10$^{-4}$, whereas 17% Eu barely exceeds ~3.5×10$^{-4}$. Thus, the numerical evidence shows that over-doping degrades PF due to the quadratic dependence on S.

Eu$^{3+}$ substitution introduces electrons and weakly hybridized 4f-derived states that raise the Fermi level in the conduction manifold, decreasing |S| (Pisarenko) and boosting σ/τ (lighter transport mass, more carriers). The same carriers elevate $\kappa_e/\tau$, while Eu-induced mass disorder is expected to lower $\kappa_L$—a favorable trade-off for thermoelectrics. Taken together, the data point to an intermediate Eu content (~8.5%) as a sweet spot for maximizing PF over a wide temperature window. Further gains are anticipated by (i) optimizing the relaxation time τ(T) through microstructural control (grain size, dislocation density), (ii) band engineering (strain or co-doping) to preserve a high S at elevated σ, and (iii) explicitly reducing $\kappa_L$ via nanostructuring while maintaining the electronic mobility.

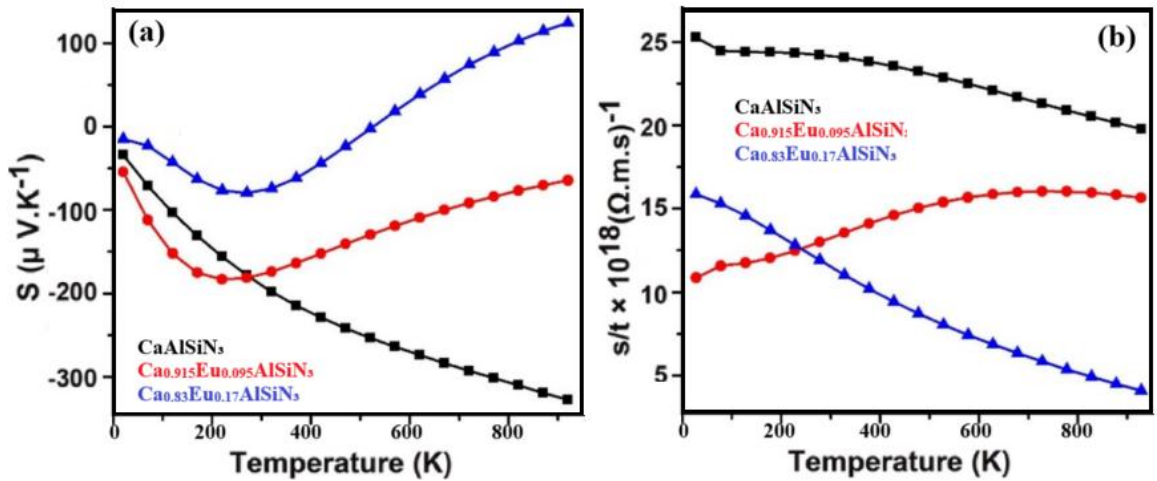

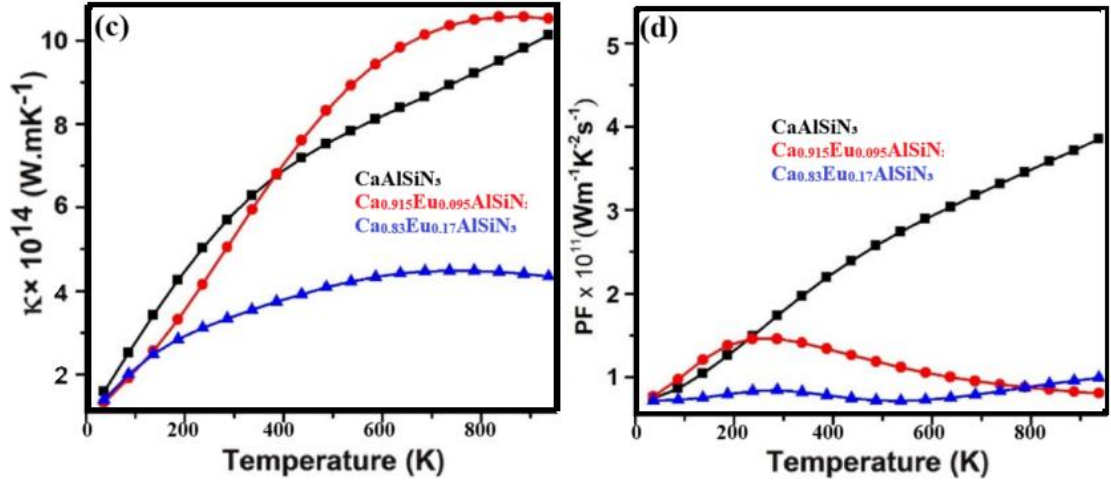

Fig. 11. (a - d) Temperature-dependent (a) Seebeck coefficient, (b) Electrical conductivity, (c) Electronic thermal conductivity, and (d) Power Factor.

## Conclusion

This work offers a comprehensive first-principles perspective on the role of $Eu^{3+}$ doping in tuning the physical properties of $CaAlSiN_3$ for photonic and lighting applications. The substitution of $Ca^{2+}$ by $Eu^{3+}$ at concentrations of 8.5% and 17% introduces discrete, highly localized 4f states within the host bandgap, facilitating red photoluminescence through parity-forbidden yet partially allowed $^5D_0 \rightarrow {}^7F\_J$ transitions. The gradual bandgap narrowing and spin polarization around Eu sites are in strong agreement with experimental PL and magnetic measurements.

The GGA+U approach accurately describes the electron correlation effects of Eu-4f orbitals, capturing the redistribution of charge density and enhanced covalency in Eu–N bonds. ELF and charge density analyses visualize this localization, reinforcing the idea of site-specific electronic tailoring. Optical calculations reveal pronounced changes in $\varepsilon_1$ and $\varepsilon_2$, with enhanced absorption and refractive index in the visible range, further supporting the material's applicability in white LED technology.

The elastic constants, bulk and shear moduli, and derived mechanical indicators (Pugh's ratio, Poisson's ratio) confirm the mechanical robustness and slight ductility of the doped systems. Phonon density of states confirms dynamical stability and provides useful thermodynamic parameters like heat capacity and vibrational entropy, which are critical for thermal management in LED packaging. Our results indicate that $Eu^{3+}$ doping around 8.5% achieves the best

<span style="color:red">thermoelectric compromise, enhancing conductivity and power factor while maintaining low lattice thermal conductivity due to enhanced phonon scattering.</span>

Together, these results confirm that Eu doping not only enhances the optoelectronic performance of $CaAlSiN_3$ but also maintains structural and thermal stability, positioning it as a superior candidate for red phosphor applications in next-generation WLEDs. The methodologies and findings in this work pave the way for rational doping strategies and computational design of efficient, thermally stable luminescent materials.

**Acknowledgement:**

The authors extend their appreciation to the Deanship of Research and Graduate Studies at King Khalid University, Kingdom of Saudi Arabia for funding this work through the Small Research Group Project under the grant number RGP.1/279/45. And This publication was also supported by the project Quantum materials for applications in sustainable technologies (QM4ST), funded as project No. CZ.02.01.01/00/22_008/0004572 by Programme Johannes Amos Commenius, call Excellent Research.

The result was developed within the project Quantum materials for applications in sustainable technologies (QM4ST), reg. no. CZ.02.01.01/00/22_008/0004572 by P JAK, call Excellent Research.

**Author Contribution:**

All authors contributed equally to the article in conceptualization, investigation, analysis, writing the original draft, review, and editing.